\documentclass{epsconf}
\usepackage{graphicx}
\usepackage{epsfig} 
\usepackage{wrapfig}
\usepackage{amsmath}
\usepackage{amssymb}

\title{Adiabatic Focusing of a Long Proton Bunch in Plasma}
\author{L.~Verra$^{1}$, E.~Gschwendtner$^{1}$, and P.~Muggli$^2$}
\institute{$^1$ CERN, Geneva 1211, Switzerland\\
$^2$ Max Planck Institute for Physics, Munich 80805, Germany}

\begin{document}
\maketitle
\section{Introduction}

\begin{figure}[!h]\centering
\includegraphics[width=80mm]{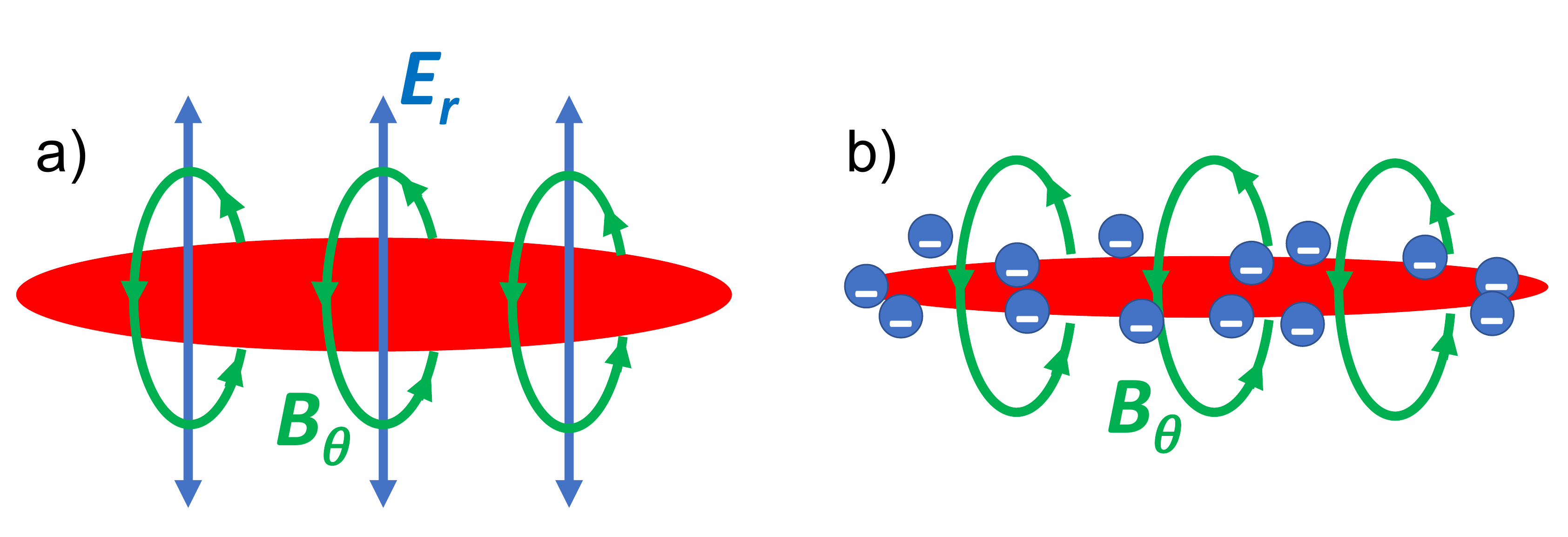}\centering
\caption{\it \small Schematic of the relativistic $p^+$ bunch (red) traveling in vacuum (a) and in plasma (b).
In (a), the total force generated by the radial defocusing electric field $E_r$ and by the azimuthal focusing magnetic field $B_{\theta}$ is close to zero ($\gamma\gg 1$).
In (b), the plasma electrons (blue circles) compensate for $E_r$: the bunch is focused by the effect of $B_{\theta}$.}
\label{fig:scheme_ad}
\end{figure}

Protons traveling in a bunch experience a repulsion due to the Coulomb force.
This force is known as space-charge force and it causes a defocusing effect on the bunch.
Since protons are moving, they also generate a magnetic field that gives rise to a focusing force.
In an azimuthal symmetric system, the electric field has essentially a radial component ($E_r$), while the magnetic field lines are circles around the bunch ($B_{\theta}$).
The total force experienced by particles in the bunch is:
$F_r = q (E_r - v_b B_{\theta}) = q E_r /\gamma^2$,
where $v_b$ is the velocity of the bunch, and $\gamma$ its Lorentz factor. 
When the bunch is relativistic ($\gamma\gg1$), the two contributions compensate for one another and the net force acting on the bunch is close to zero (see Figure~\ref{fig:scheme_ad}(a)). 

\par When the bunch travels in a neutral plasma, plasma electrons migrate so as to neutralize the space-charge field of the bunch~\cite{PWFA}. 
As a consequence, the focusing force generated by the azimuthal magnetic field of the bunch is no longer balanced by the defocusing force due to the space charge electric field, and the bunch is therefore focused~\cite{JAPAN,UCLA_ADIABATIC,BERKLEY_ADIABATIC} (see Figure~\ref{fig:scheme_ad}(b)).

When the bunch-plasma system evolves over successive equilibrium states, the process is adiabatic and there is no oscillation of the beam envelope nor of the plasma electron motion. 
However, when the plasma electrons move so fast that they overshoot, the perturbation of the plasma electron density provides a restoring force that induces an oscillation of the plasma electrons with
angular frequency equal to the plasma electron frequency $\omega_{pe}=\sqrt{\frac{n_{pe}e^2}{m_e \epsilon_0}}$ ($n_{pe}$ is the plasma electron density, $m_e$ is the mass rest of the electron, $e$ is the elementary charge and $\varepsilon_0$ is the vacuum permittivity).
The local charge non-neutrality sustains wakefields that can be used for particle acceleration and that can modify the bunch itself, adding to the self-focusing force, and lead to self-modulation (SM)~\cite{KARL:PRL, MARLENE:PRL}.

\par The Advance WAKefield Experiment (AWAKE) at CERN~\cite{PATRIC:READINESS} uses a long relativistic proton ($p^+$) bunch driving wakefields in plasma with $n_{pe}=\mathcal{O}(10^{14})\,$cm$^{-3}$ to accelerate externally injected electrons to GeV energies~\cite{NATURE}. 
Since the $p^+$ bunch duration $\sigma_t$ is much longer than the plasma electron period $T_{pe}=2\pi/\omega_{pe}$, the bunch self-modulates into a train of microbunches that can drive wakefields resonantly~\cite{KARL:PRL}.
The overall transverse size along the bunch appears to increase because of protons that are defocused in the SM process~\cite{MARLENE:PRL}.
However, we observe in experiments that the front of the bunch undergoes a global focusing effect due to the adiabatic response of the plasma~\cite{LIVIO}.
To isolate the adiabatic focusing effect from the SM effect, we lower the plasma electron density so that SM does not take place and the bunch is affected only by the self-focusing force.

\par For a bunch with geometric emittance $\epsilon_g$, self-focused by its own azimuthal magnetic field, the transverse size $\sigma_r$ evolves according to the envelope equation: $\sigma_r''=\frac{\epsilon_g}{\sigma_r^3} - \frac{K^2}{\sigma_r}$.
The strength of the focusing force $K$ is defined as:
$K(t)=\sqrt{\frac{ e^2 \lambda_b(t)}{2\pi \gamma m_p c^2}(1-e^{-1/2})}$, where $\lambda_b(t)$ is the bunch charge density per unit length as a function of time $t$ along the bunch, $m_p$ is the rest mass of the proton and $c$ is the speed of light.
When the bunch is Gaussian, $\lambda_b(t)$ varies along the bunch: it is maximum at the center ($t=0$) and symmetric with respect to its center.
In the longitudinal slices where the focusing force is strong enough, the transverse distribution of the bunch may reach an equilibrium size at $\sigma_r(t)=\epsilon_g/K(t)$ before the plasma exit.
The size then remains constant until the plasma exit.

\section{Experimental Results}
In AWAKE, the plasma is generated by ionizing rubidium with a relativistic ionization front in a $10$-m-long vapor source.
The $p^+$ bunch with normalized emittance $\epsilon_N=\frac{v_b}{c} \gamma \epsilon_g=2\,$mm mrad is focused at the plasma entrance to a transverse size $\sigma_{r0}=0.2\,$mm.
After the plasma exit, the beam propagates in vacuum and, after $3.5\,$m, crosses a screen emitting optical transition radiation (OTR).
The OTR is transported and imaged onto the entrance slit of a streak camera, that produces time-resolved images of the $p^+$ bunch charge density distribution near the bunch axis. 

\par The SM process is due to the transverse wakefields driven by the bunch, acting back on the bunch itself and initiating a positive feedback loop. 
Since the wakefields are periodically focusing and defocusing, some protons are squeezed towards the axis into microbunches~\cite{KARL:PRL}, while others are defocused out of the wakefields~\cite{MARLENE:PRL}. 
As a result, the overall transverse extent of the bunch distribution (measured with the streak camera) where SM dominates is larger than in the case without plasma and increases along the bunch~\cite{LIVIO}.
However, the front of the bunch (where growth of SM is small) is focused by the adiabatic response of the plasma, and the transverse extent of the bunch at the screen follows the same trend both with and without seeding~\cite{LIVIO}.

%
\begin{figure}[!h]\centering
\includegraphics[width=120mm]{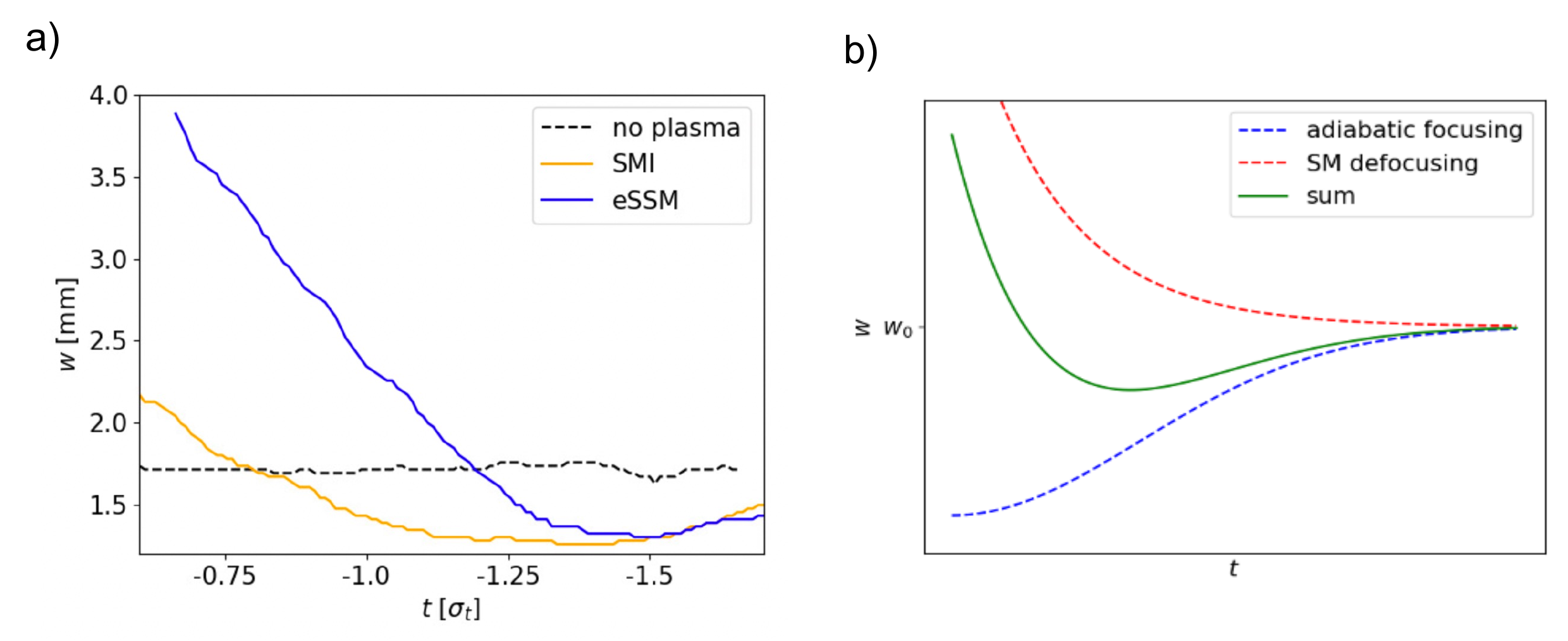}\centering
\caption{\it \small 
Transverse extent $w$ along the bunch (green line, the bunch travels from left to right) calculated as the sum of the contribution from the adiabatic focusing (blue dashed line) and from SM defocusing (red dashed line).}
\label{fig:extent}
\end{figure}

\par Figure~\ref{fig:extent}(a) shows the transverse extent $w$ along the bunch as measured experimentally after propagation in vacuum (black dashed line) and after propagation in plasma ($n_{pe}=0.97\cdot 10^{14}\,$cm$^{-3}$) in case of no seeding (SM occurs as an instability: SMI, orange line) and in case of seeding with a preceding electron bunch (SM occurs in a controlled and reproducible way: eSSM~\cite{LIVIO}, blue line).
In both cases with plasma, the transverse extent at the front of the bunch is smaller than without plasma and follows the same trend due to the self-focusing effect.
This shows that this effect due to the plasma adiabatic response dominates over the duration in front of where the amplitude of the wakefields remains small.
It occurs in a similar way independently of the seed wakefields.
Later along the bunch, the defocusing effect due to SM (that grows along the bunch and along the plasma) is dominant, and the transverse extent reaches larger values than without plasma. 
The transition from global (adiabatic) focusing to global (SM) defocusing depends on the amplitude of the seed wakefields and on the growth of the instability~\cite{LIVIO}.

\par The superposition of the two effects on the transverse extent of the bunch is schematically shown in Figure~\ref{fig:extent}(b). 
The blue dashed line shows the effect of focusing on the transverse extent $w$ of the adiabatic response of the plasma (no SM), that increases along the bunch following its Gaussian density distribution ($\propto e^{t^2/\sigma_t^2}$). 
The red dashed line shows the defocusing effect due to SM, that grows along the bunch. 
The sum of the two components (green line) shows first a decrease in $w$ to values smaller than the initial one ($w_0$) and then an increase with the growth of SM, leading to a qualitative agreement with the experimental results shown in Figure~\ref{fig:extent}(a).

\par Preliminary experimental results indicate that when the $p^+$ bunch with same parameters as in Figure~\ref{fig:extent}(a) travels in plasma with much lower plasma electron density ($n_{pe}\lesssim10^{12}\,$cm$^{-3}$), SM does not develop, but the transverse extent at the bunch front follows the same trend as with higher plasma electron densities. 
This result shows that the strength of the adiabatic focusing is independent of $n_{pe}$, as long as there are enough plasma electrons in each longitudinal slice to compensate for the space-charge field of the bunch. 
Moreover, we observe that when decreasing the charge density of the $p^+$ bunch that propagates in low-density plasma, SM does not occur unless seeded, indicating a transition between the adiabatic and the wakefields regime.
\section{Conclusions}

We discussed the principles of the adiabatic response of the plasma to the presence of a relativistic charged particle bunch. 
Since the free plasma electrons move to compensate for the space-charge electric field of the bunch, it is focused by its own magnetic field.
Each longitudinal slice is focused independently from the rest of the bunch, and where the focusing is sufficiently strong the size can reach an equilibrium value before the end of the plasma.
This focusing effect is visible at the bunch front even when SM occurs later along the bunch.
The transition to the global defocusing effect depends on the growth of SM.
In AWAKE, experiments are ongoing to further study the adiabatic focusing of the long $p^+$ bunch in plasma.

\end{document}